\documentclass[12pt]{article}
\usepackage{amsfonts,amssymb,amsmath,amscd}

\newcommand{\ra}{\rightarrow}
\newcommand{\Tr}{{\rm Tr}}
\newcommand{\gst}{g_{\rm st}}
\newcommand{\A}{{\mathcal A}}
\newcommand{\hc}{{\hat c}}
\newcommand{\tmu}{{\tilde \mu}}
\newcommand{\ZZ}{{\mathbb Z}}

\title{Noncritical Superstrings in a Ramond-Ramond Background}
\author{Anton Kapustin\\{\it California Instutute of Technology, Pasadena, CA 91125, U.S.A.}}

\begin{document}

\begin{titlepage}

\maketitle

\begin{abstract}
We use the recently found matrix description of noncritical superstring theory of Type 0A to
compute tachyon scattering amplitudes in a background with a RR flux. 
We find that after the string coupling is multiplicatively
renormalized, the amplitudes in any genus become polynomial in the RR flux.
We propose that in the limit where both the string coupling and the RR flux go to infinity,
the theory has a weakly-coupled description
in terms of another superstring theory with a vanishingly small RR flux. This duality
exchanges the inverse string coupling and the 0-brane charge. The dual superstring
theory must have a peculiar property that its only field-theoretic degree of freedom
is a massless RR scalar.

\end{abstract}

\end{titlepage}

\section{Introduction}

Understanding superstring theory in a Ramond-Ramond background is a long-standing
problem. In the NSR formalism, Ramond-Ramond vertex operators are twist operators for world-sheet fields,
and it is far from clear what ``exponentiating'' them means. An additional problem is that insertions of RR
vertex operators makes the supercurrent multi-valued, which seems to imply that the world-sheet theory of a RR
background is not superconformal. On the other hand, the Green-Schwarz (GS) formalism is
manageable only in the light-cone gauge, and this puts serious limitations on the kinds of backgrounds
one can study. A variety of approaches to this problem have been studied in the literature; for recent
progress see~\cite{B1,B2,BM,B3} and references therein.

Noncritical superstrings are an interesting laboratory for studying RR backgrounds, because one can hope
to find simple RR backgrounds which lead to an integrable theory. Recently, one such background has been
constructed, and the corresponding large-$N$ matrix description has been identified~\cite{hat} (see also
Ref.~\cite{TT}). It is a $\hc=1$ superstring
of Type 0A with a RR 2-form field-strength $F_{(2)}$ turned on. The $F$-flux $m$ measures the 0-brane charge and
is quantized. The dual matrix model is proposed to be a gauged matrix model with gauge group $U(N)\times U(N+m)$
and a single complex matter field $t$ transforming in the bi-fundamental representation. In the double-scaling
limit the potential can be taken to be 
$$
V(t)=-\Tr(t^\dag t).
$$
This theory is equivalent
to a theory of free fermions on a plane, all of which have angular momentum $m,$ and moving in an external potential
$$
V(\vec{r})=-\frac{r^2}{2}.
$$
Since the angular momentum is fixed, each fermion can be regarded as moving on a half-line $r\geq 0$
in an effective potential
\begin{equation}\label{Veff1}
V_{eff}(r)=-\frac{r^2}{2}+\frac{m^2-\frac{1}{4}}{2r^2}.
\end{equation}
For $m\neq 0$ the condition that the wave-function is square-integrable on a plane and non-singular at the origin
is equivalent to the condition that the wave-function on the half-line be square-integrable. For $m=0$
there is an additional condition: the limit
$$
\underset{r\ra 0}{\lim} \frac{\psi(r)}{\sqrt r}
$$
must be finite. Type 0A superstring is recovered in the limit where the Planck constant and the Fermi-level go to
zero, with their ratio fixed.

Once the problem is reduced to free fermions on a half-line moving in an external potential, we can use
the methods of Ref.~\cite{Smatrix} to compute non-perturbative scattering amplitudes for ``tachyons''.
The only ingredient needed is the reflection amplitude for the particular potential we are considering.
In fact, the potential of the above kind has already been considered in the string literature in connection
with the so-called deformed matrix model~\cite{deformed}, and the reflection factor has been computed~\cite{DKR}. 
The deformed matrix
model is a matrix quantum mechanics with the potential 
$$
V(X)=\Tr\left(- X^2 +\frac{M}{X^2}\right),
$$ 
where $X$ is a Hermitian matrix, and $M$ is a real number.
It obviously reduces to free fermions on a half-line in an external potential  Eq.~(\ref{Veff1}), provided
we set $M=m^2-\frac{1}{4}.$
So all we have to do to compute scattering amplitudes in a noncritical RR background is to put 
together~\cite{Smatrix} and \cite{DKR}. 

\section{Weak-field scattering amplitudes}

Genus expansion in string theory corresponds to expansion of the free fermion correlators in powers of $1/\mu^2$, 
where $\mu$ is the Fermi-level. We must also decide how the parameter $m$ scales with $\mu$.
In this section we keep $m$ fixed, while in the next section we will let $m=f\mu$. To understand better
the meaning of these two scalings, recall that the most natural normalization for the RR fields is such that their
kinetic term in the low-energy action is independent of the dilaton. With this normalization, the Bianchi 
identity for a RR field-strength $F$ takes the usual form $dF=0$, and the $F$-flux is quantized. We will call this
``target-space normalization.''
String perturbation theory, on the other hand, prefers a differently normalized
field-strength $F'$, which has the property that its kinetic term has the usual factor $e^{-2\Phi}$, where $\Phi$ is the
dilaton~\cite{Polch}. We will call this ``world-sheet normalization.'' The field $F'$ satisfies a modified Bianchi identity
$$
dF'=d\Phi\wedge F',
$$
The corresponding RR potentials are related by
$$
C=e^{-\Phi}C'.
$$
The parameter $m$ is the flux of $F$.Keeping it fixed in the limit $\gst\sim \mu^{-1}\ra 0$ means that $F'$ is of order $g_{st}$.
Therefore we will call this the weak-field regime. In this regime at any order in $\gst$ we need to take into account
only a finite number of insertions of RR vertex operators, i.e. the RR flux is treated perturbatively.
On the other hand, if $f=m/\mu$ is kept fixed, then $F'$ is of order $1$, and already at tree level one has to allow
arbitrary number of RR insertions. This will be referred to as the strong-field regime.

The double-scaled effective potential for fermions on a half-line is
$$
V(r)=-\frac{r^2}{2}+\frac{m^2-\frac{1}{4}}{2r^2}.
$$
The corresponding Euclidean reflection amplitude is~\cite{DKR}:
\begin{equation}\label{reflamp}
R_q=\left(\frac{4}{m^2+\mu^2-\frac{1}{4}}\right)^{\frac{|q|}{2}}
\frac{\Gamma\left(\frac{1}{2}\left(1+|m|+|q|-i\mu\right)\right)}{\Gamma\left(\frac{1}{2}\left(1+|m|-|q|+i\mu\right)\right)}.
\end{equation}
Here $\mu$ is the Fermi level and $q$ is the Euclidean momentum. We also omitted an unimportant $q$-independent phase.
The perturbative limit corresponds to $\mu\ra\infty$, and one identifies $\gst\sim \mu^{-1}$.

The reflection factor is invariant with respect to $m\ra -m$. In the rest of the paper we assume
that $m\geq 0$. 

The rules for computing non-perturbative scattering amplitudes for the collective field are the same as in Ref.~\cite{Smatrix}. 
For example,
the 2-point function and 3-point function are given by
\begin{align}
\A_2(q,-q)&=\int_0^q R_{q-x} R^*_x dx,\\
\A_3(q_1,q_2;-q)&=-i\left(\int_{q_1}^q R_{q-x} R^*_x dx+\int_{q_2}^q R_{q-x} R^*_x dx - \int_0^q R_{q-x} R^*_x dx\right),
\end{align}
where $q=q_1+q_2$, and $q,q_1,q_2$ are all positive. For the 4-point function there are several distinct kinematical 
configurations which correspond to different expressions. For example, if $q_1,q_2,q_3>0$ and $q_4=-q<0$, we have
\begin{align}
\A_4(q_1,q_2,q_3;-q)&=&\int_{q_1}^q+\int_{q_2}^q+\int_{q_3}^q -\int_{q_1+q_2}^q-\int_{q_1+q_3}^q-\int_{q_2+q_3}^q-\int_0^q 
R_{q-x} R^*_x dx.
%%\A_4(q_1,q_2;-q_3,-q_4)&=&-\left(\int_{q_1}^{q_1+q_2}+\int_0^{q_2}\right) R_{q_1+q_2-x} R^*_x dx\\
%%+& \frac{1}{2}\left\{\int_0^{q_2} R_{q_3-x} R_{q_2-x} R^*_{x+q1-q3} R^*_x dx+\int_{q_3-q_2}^q_3 R_{q_3-x} R_{q_1-x} 
%%R^*_{x+q_2-q_3} R^*_x+\left(q_3\ra q_4\right)\right\}.
\end{align}

Expanding the reflection amplitude in powers of $1/\mu$, we obtain the genus expansion for the tachyon scattering 
amplitudes.\footnote{Scattering amplitudes computed using continuum methods differ from the collective field scattering
amplitudes by so-called leg factors which depend on momenta~\cite{DiFKu}. 
To bring continuum results in agreement with the matrix model,
one has to absorb these factors into the tachyon vertex operators. This means that the tachyon field and the collective
field of the matrix model are related by a linear but non-local transformation. In this paper we always assume
that such a redefinition of the tachyon vertex operators has been performed.}
The 2-point function up to order $1/\mu^4$ is given by
\begin{multline}
\A_2(q;-q)=q^2\left(\frac{1}{q}-\frac{(q-1)(q^2-q-1)}{24}\mu^{-2}+\frac{1}{5760}\left(3q^7-28q^6+88q^5 
\right.\right.\\
\left.\left.-86 q^4-180q^3+480q^2+5q-582+240 m^2 (2q^3-6q^2+9)\right)\mu^{-4}+
O(\mu^{-6})\right).
\end{multline}
The 3-point function up to 1-loop order is given by
\begin{multline}
\A_3(q_1,q_2;-q)=\frac{q_1 q_3 q_3}{\mu}\left(1-\frac{1}{24}\left(q^4-2q^3(q_1+2)+2q^2(q_1^2+3q_1+2)\right.\right.\\
\left.\left. -q(6q_1^2+4q_1-1)+4(q_1^2-2)+24 m^2\right)\mu^{-2}+O(\mu^{-4})\right).
\end{multline}
Here $q_3=-q=-(q_1+q_2).$
The tree-level 4-point function is given by
\begin{equation}
\A_4(q_1,q_2,q_3,q_4)=\mu^{-2}\prod_{i=1}^4 |q_i| \left({\rm max}\{|q_j|\}-1+O(\mu^{-2})\right).
\end{equation}
We do not give here a somewhat lengthy expression for the contribution to the 4-point function of relative order
$1/\mu^2$, but remark that it is polynomial in momenta and quadratic in $m$. We also checked that the leading
contribution to the 5-point function in a particular kinematic configuration ($4\ra 1$ scattering) is independent
of $m$.

Recall that noncritical superstring scattering amplitudes in the absence of RR flux differ from their
critical counterparts in at least two major ways. First, they are not analytic, or even differentiable, functions of momenta. 
Instead, the phase space (i.e. the space of momenta) is divided into ``wedges'', and the scattering amplitudes are given by 
different analytic expressions in each wedge. Second, these analytic expressions are polynomials in momenta. From the above 
formulas we see that turning on the RR flux preserves these qualitative features of the scattering amplitudes.
Note also that at any order in $1/\mu$ expansion the dependence on $m$ is polynomial, and leading-order
expressions are independent of $m$. As discussed above, this agrees with the identification of $m$ as the flux of the 
RR field $F$.

\section{Strong-field scattering amplitudes}

We now set $m=f\mu$ and re-expand the scattering amplitudes in powers of $1/\mu$. Apriori, one could
expect arbitrarily complicated powers of $f$ and momenta at any order in $1/\mu$ expansion. We will see below that
this does not happen. The dependence on momenta remains polynomial, and, even more surprisingly, the dependence on the
RR flux is very simple as well. Namely, apart from a multiplicative renormalization of the string coupling by a function
of $f$, the dependence on $f$ is polynomial. 

Let us now present the scattering amplitudes in the strong-field regime. The 2-point function up to 2-loop order
is given by
\begin{multline}
\A_2(q;-q)=q^2\left(\frac{1}{q}-\frac{(q-1)(q^2-q-1)+f^2(2q^2-7)}{24(1+f^2)^2}\mu^{-2}\right.\\
\left.+\frac{\mu^{-4}}{5760(1+f^2)^4}\left(3q^7-28q^6+88q^5-86q^4-180q^3+480q^2+5q-582\right.\right.\\
\left.\left.+f^2(28q^6-240q^5+506q^4+640q^3-2880q^2-70q+4212)\right.\right.\\
\left.\left. +f^4(24q^5-96q^4-140q^3+480q^2+245q-582)\right)+O(\mu^{-6})\right).
\end{multline}
The 3-point function up to 1-loop order is given by
\begin{multline}
\A_3(q_1,q_2,q_3)=\frac{q_1 q_2 q_3}{(1+f^2)\mu}\left(1-\frac{\mu^{-2}}{24(1+f^2)^2}\left(q^4-2q^3(q_1+2)\right.\right.\\
\left.\left.+2q^2(q_1^2+3q_1+2)-q(6q_1^2+4q_1-1)+4(q_1^2-2)\right.\right.\\
\left.\left.+f^2(4q^3-6q^2(q_1+2)+q(6q_1^2+12q_1-7)-12(q_1^2-2))\right)+O(\mu^{-4})\right).
\end{multline}
The tree-level 4-point function in the kinematic configuration corresponding to $3\ra 1$ scattering is given by
$$
\A_4(q_1,q_2,q_3;-q)=\frac{\mu^{-2}}{(1+f^2)^2}\prod_{i=1}^4 |q_i|\left(q-1+f^2\right).
$$
We remind that $q=q_1+q_2+q_3=-q_4$. 

From the above formulas we see that after multiplicative renormalization
$$
\mu\ra\tmu=(1+f^2)\mu,
$$
the scattering amplitudes at each order in $1/\tmu$ expansion become polynomials in $f$. The degree of the polynomial
grows with the order of the genus expansion. By examining the reflection amplitude $R_q$, one can show
that these properties hold for arbitrary $n$-point functions and to all orders in $1/\mu$ expansion. 

We also note that tree level 2- and 3-point functions do not depend on $f$ at all, if $\tmu$ is kept fixed. 
But higher-point functions at tree level do depend on $f$ even after renormalization of the string coupling.
These simple results looks mysterious from the viewpoint of world-sheet theory.

\section{A peculiar duality}

The free-fermion theory with the potential~Eq.(\ref{Veff1}) admits another ``perturbative'' limit, where $\mu$ is kept fixed and
$m$ is taken to infinity. This was first noted in the context of the deformed matrix model~\cite{deformed}. The idea behind
this limit is the following. The usual perturbative limit is based on the fact that correlation functions are
singular (non-analytic) at the point where the Fermi energy coincides with the top of the quadratic potential.
Finite Fermi energy $\mu$ regularizes this singularity, and string perturbation theory picks out the non-analytic
dependence on $\mu$ near $\mu=0$. One can also regularize the singularity by adding the $1/r^2$ piece
to the potential. The new perturbative limit is obtained by picking out terms non-analytic in $M$, i.e. by
performing $1/M$ expansion. The Fermi level $\mu$ can be arbitrary; one can even set it to zero.
Qualitatively, this scaling limit looks very much like the usual one, so one may hypothesize that it corresponds to
some perturbative noncritical string theory. The role of string coupling is played by $M^{-1/2}$.
Attempts to identify this string theory have been unsuccessful so far.\footnote{In Ref.~\cite{deformed} it was
conjectured that the deformed matrix model describes the 2d black hole background for the bosonic string, but in our 
opinion this is unlikely.} Various scattering amplitudes have been computed in Refs.~\cite{deformed,DR,Dan,DKR}.

In this section we make some observation which may point to the correct string theory for this limit.
From the point of view of the Type 0A theory, the limit $\mu=const$, $m\ra\infty$ is a limit where the RR flux (in the
target-space normalization) is taken to infinity with the string coupling $\gst\sim \mu^{-1}$ fixed. In fact, one can 
even set $\mu=0$, and the limit $m\ra\infty$ is still
well-defined. It is natural to assume that the dual theory is again a superstring theory. Further, the only 
field-theoretic degree of freedom is still
the Fermi-level (i.e. the collective field), so we must be dealing with a superstring in $d=1$. 

Note that the $\mu$ and $m$ enter the reflection amplitude Eq.~(\ref{reflamp}) in a similar way. 
Since in the limit $m\ra\infty$ the
parameter $1/m$ becomes the dual string coupling, it is reasonable to suspect that in this limit $\mu$ becomes
the dual RR flux $m'$ (in the target-space normalization). If we define $\gst'=1/m$, $m'=\mu$, and $f'=\gst' m'$, 
then the relation between the parameters 
of the two dual descriptions is
\begin{equation}\label{duality}
\gst'=\frac{\gst}{f},\qquad f'=\frac{1}{f}.
\end{equation}
In particular, the dual string background with $f'=0$ and finite string coupling corresponds to
the limit $\gst\ra\infty,f\ra \infty$ with $\gst/f$ kept fixed. We will see below that scattering amplitudes
written in terms of $\gst'$ and $f'$ have the structure consistent with their interpretation as the dual string
coupling and the dual RR flux (in the world-sheet normalization). Note that the transformation Eq.~(\ref{duality}) 
is reminiscent of T-duality along a circle, where the role of radius is played by the RR flux $f$.

The authors of Ref.~\cite{hat} proposed to study a related limit 
$$
m\ra\infty,\quad \mu\ra\infty, \quad \frac{\mu}{m}=const.
$$
From our perspective, this is the zero-coupling/strong-field regime of the dual theory, i.e. one takes the dual
string coupling $\gst'$ to zero while keeping the dual RR flux $f'$ fixed.

The interpretation of $\mu$ as the RR flux in the dual theory explains the following interesting 
observation about the deformed matrix model made in Refs.~\cite{deformed,DKR}.
If one sets $\mu=0$, then all $n$-point functions in the deformed matrix model with odd $n$ vanish 
(to all orders in $1/m$ expansion).
It is easy to check that for $\mu\neq 0$ $n$-point functions with odd $n$ are odd functions of $\mu$,
while $n$-point functions with even $n$ are even functions of $\mu$. For example, the 2-point function up to 2-loop order is
\begin{multline}
\A_2(q;-q)=q^2\left(\frac{1}{q}-\frac{2q^2-7}{24}m^{-2}+\frac{1}{5760}\left(24q^5-96q^4-140q^3+480q^2 \right.\right.\\
\left.\left. +245q-582-240\mu^2(q^3-6q^2+15)\right)m^{-4}+O(m^{-6})\right),
\end{multline}
while the 3-point function up to 1-loop order is
\begin{multline}
\A_3(q_1,q_2;-q)=\mu m^{-2} q_1q_2q_3\left(1-\frac{1}{24}\left(4q^3-6q^2(q_1+2)+q(6q_1^2+12q_1-7)\right.\right.\\
\left.\left. -12(q_1^2-2)+24\mu^2\right)m^{-2}+O(m^{-4})\right).
\end{multline}
Let us also write down the leading-order 4-point function:
\begin{equation}\label{fourpt}
\A_4(q_1,q_2,q_3,q_4):=m^{-2}\prod_{i} |q_i|\left(1+O(m^{-2})\right).
\end{equation}
Note that the leading term in the 3-point function scales as $\mu m^{-2}$. Since we interpreted $m^{-1}$ as $\gst'$,
this seems like a wrong scaling for a 3-point function on a sphere. However, if we recall that $\mu$ also scales with
$m$, $\mu=f'm$, we see that the leading piece is, in fact, of order $\gst'$. The 4-point function is of order $\gst'^2$,
as expected, and is an even function of $\mu$.

These properties of the amplitudes can be summarized by saying that the system admits a $\ZZ_2$ symmetry which 
reverses the signs of both $\mu$ and the collective field. If $\mu$ is
interpreted as a RR flux, it is tempting to identify this $\ZZ_2$ symmetry as $(-1)^{F_L},$ where $F_L$ is
the left-moving target-space fermion number. Then the collective
field must be a RR scalar. Thus we get the following important hint about the dual superstring theory: its only
field-theoretic degree of freedom is a massless RR scalar, while the usual NS-NS ``tachyon'' is absent.

So far our only evidence that the limit $m\ra\infty$ is described by a dual weakly-coupled superstring theory has been the
structure of scattering amplitudes for the collective field. Namely, the formulas for the $n$-point
functions in this limit look similar to those in the weak-coupling limit $\mu\ra\infty$, if we exchange
$m$ and $\mu$. Another interesting test is to compactify the
Euclidean time on a circle and check if the matrix-model partition function enjoys T-duality. Since
$\mu$ is supposed to be the dual RR flux, we expect to get a self-dual theory only for $\mu=0$.
Properties of the deformed matrix model partition function for $\mu=0$ have been discussed in Refs.~\cite{DR} and~\cite{Dan2}.
In Ref.~\cite{DR} it was found that fixed-genus partition functions do not behave nicely under T-duality. In that paper
it was assumed that the genus-counting parameter is $M^{-1}$, which is related to our $m$ by
$$
M=m^2-\frac{1}{4}.
$$
However, it was noticed in Ref.~\cite{Dan2} that if one takes 
$$
\frac{4}{1+4M}=\frac{4}{m^2}
$$
as the genus-counting parameter, then T-duality is restored. This observation provides support for our conjecture that 
the limit $m\ra\infty$ is described by a weakly-coupled superstring theory with string coupling $\gst'\sim m^{-1}$.

\section{Discussion}

We have computed collective-field scattering amplitudes in the noncritical Type 0A superstring theory with Ramond-Ramond flux. 
Weak-field expansion of the amplitudes is in agreement with expectations from string perturbation theory. Unexpectedly,
scattering amplitudes at large RR flux and fixed genus turned out to have a very simple structure: after a multiplicative
renormalization of the string coupling by a simple function of the flux $f=m \gst$, all amplitudes are polynomial in $f$.
It would interesting to understand this behavior from a world-sheet point of view.

We also argued that in the limit $f\ra\infty,\gst\ra\infty$ with $\gst/f$ fixed there is a 
weakly-coupled description in terms of another superstring theory with zero RR flux. More generally, the relation between
the parameters of the original and dual string theories is given by Eq.~(\ref{duality}). The dual theory 
has the property that its only field-theoretic degree of freedom is a massless RR scalar. 

Let us make a few comments about the world-sheet description of the dual superstring theory. It must be a superstring
theory in $1+1$ dimensions with a time-translation invariance. The only known background of this kind is the $N=1$
Liouville theory coupled to $\hc=1$ matter. Therefore the simplest guess for the dual world-sheet theory is some
projection of the standard $\hc=1$ background. But, the super-Liouville interaction precludes any chiral GSO projection,
while the non-chiral GSO projection gives the two Type 0 theories
discussed in Refs.~\cite{hat,TT}. Thus one must look for more exotic possibilities. 

\section*{Acknowledgments}

I am grateful to Jaume Gomis for useful discussions. This work was supported in part 
by the DOE grant DE-FG03-92-ER40701.


\begin{thebibliography}{99}

\bibitem{B1} N.~Berkovits, ``Super-Poincare covariant quantization of the superstring,''
JHEP {\bf 0004}, 018 (2000) [arXiv:hep-th/0001035].

\bibitem{B2} N.~Berkovits, ``Covariant quantization of the superstring,''
Int.\ J.\ Mod.\ Phys.\ A {\bf 16}, 801 (2001)
[arXiv:hep-th/0008145].

\bibitem{BM} N.~Berkovits and J.~Maldacena, ``N = 2 superconformal description of superstring in Ramond-Ramond plane 
wave backgrounds,'' JHEP {\bf 0210}, 059 (2002) [arXiv:hep-th/0208092].

\bibitem{B3} N.~Berkovits, ``N = 2 sigma models for Ramond-Ramond backgrounds,''
JHEP {\bf 0210}, 071 (2002) [arXiv:hep-th/0210078].


\bibitem{hat} M.~R.~Douglas, I.~R.~Klebanov, D.~Kutasov, J.~Maldacena, E.~Martinec and N.~Seiberg,
``A new hat for the c = 1 matrix model,'' arXiv:hep-th/0307195.

\bibitem{TT} T.~Takayanagi and N.~Toumbas,
``A matrix model dual of type 0B string theory in two dimensions,''
JHEP {\bf 0307}, 064 (2003) [arXiv:hep-th/0307083].


\bibitem{Smatrix} G.~W.~Moore, M.~R.~Plesser and S.~Ramgoolam,
``Exact S matrix for 2-D string theory,'' Nucl.\ Phys.\ B {\bf 377}, 143 (1992)
[arXiv:hep-th/9111035].

\bibitem{deformed} A.~Jevicki and T.~Yoneya,
``A Deformed matrix model and the black hole background in two-dimensional string theory,''
Nucl.\ Phys.\ B {\bf 411}, 64 (1994) [arXiv:hep-th/9305109].

\bibitem{DKR} K.~Demeterfi, I.~R.~Klebanov and J.~P.~Rodrigues,
``The Exact S matrix of the deformed c = 1 matrix model,''
Phys.\ Rev.\ Lett.\  {\bf 71}, 3409 (1993) [arXiv:hep-th/9308036].

\bibitem{Polch} J.~Polchinski, ``String Theory. Vol. 2: Superstring Theory And Beyond,'' Cambridge University Press (1998).

\bibitem{DiFKu} P.~Di Francesco and D.~Kutasov,
``World-sheet and space-time physics in two-dimensional (super)string theory,''
Nucl.\ Phys.\ B {\bf 375}, 119 (1992) [arXiv:hep-th/9109005].

\bibitem{DR} K.~Demeterfi and J.~P.~Rodrigues, ``States and quantum effects in the collective field theory of a 
deformed matrix model,'' Nucl.\ Phys.\ B {\bf 415}, 3 (1994) [arXiv:hep-th/9306141].

\bibitem{Dan} U.~H.~Danielsson, ``A Matrix model black hole,''
Nucl.\ Phys.\ B {\bf 410}, 395 (1993) [arXiv:hep-th/9306063].

\bibitem{Dan2} U.~H.~Danielsson, ``The deformed matrix model at finite radius and a new duality symmetry,''
Phys.\ Lett.\ B {\bf 325}, 33 (1994) [arXiv:hep-th/9309157].




\end{thebibliography}
\end{document}